\documentclass[apj]{emulateapj}\usepackage{apjfonts}\usepackage{aas_macros}
\usepackage{natbib}
\usepackage{amsmath}
\usepackage{amssymb}
\usepackage{color}
\bibpunct[, ]{(}{)}{,}{a}{}{,}

\newcommand{\pct}{\,\%}
\newcommand{\dif}{\mathrm{d}}
\newcommand{\diff}[2]{\frac{\dif#1}{\dif#2}}
\newcommand{\pma}[2]{{}^{+#1}_{-#2}}
\newcommand{\lgd}{\log_{10}}
\newcommand{\lgm}{\log_{10} m/\msun}
\newcommand{\xil}{\xi_\mathrm{L}}
\newcommand{\xibd}{\xi_\mathrm{BD}}

\newcommand{\ximod}{\xi_\mathrm{theo}}
\newcommand{\xiobs}{\xi_\mathrm{obs}}
\newcommand{\xires}{\xi_\mathrm{res}}

\newcommand{\rpop}{\mathcal{R}_\mathrm{pop}}

\newcommand{\nbod}{N_\mathrm{bod}}
\newcommand{\nsng}{N_\mathrm{sng}}
\newcommand{\nbny}{N_\mathrm{bin}}
\newcommand{\nsys}{N_\mathrm{sys}}
\newcommand{\msun}{M_{\sun}}
\newcommand{\tmsun}{\mbox{$\msun$}}

\newcommand{\mmax}{m_\mathrm{max}}
\newcommand{\mecl}{M_\mathrm{ecl}}
\newcommand{\ppair}{p_\mathrm{pair}}
\newcommand{\mpri}{m_\mathrm{prim}}
\newcommand{\msec}{m_\mathrm{comp}}
\newcommand{\fq}{\gamma}

\newcommand{\fbin}{f}

\newcommand{\rhm}{\mbox{$r_h$}}

\newcommand{\oversim}[2]{\protect{\mbox{\lower0.5ex\vbox{%
   \baselineskip=0pt\lineskip=0.2ex
   \ialign{$\mathsurround=0pt #1\hfil##\hfil$\crcr#2\crcr\sim\crcr}}}}} 
\newcommand{\simless} {\mbox{$\,\mathrel{\mathpalette\oversim<}\,$}} 
\begin{document}
\title{CHARACTERIZING THE BROWN DWARF FORMATION CHANNELS FROM THE INITIAL MASS FUNCTION AND BINARY-STAR DYNAMICS}
\author{Ingo Thies\altaffilmark{1}, Jan Pflamm-Altenburg\altaffilmark{1}, Pavel Kroupa\altaffilmark{1}, Michael Marks\altaffilmark{1}}
\altaffiltext{1}{Helmholtz-Institut f\"ur Strahlen- und Kernphysik (HISKP), Universit\"at Bonn, Nussallee 14--16, D-53115, Germany}
\begin{abstract}
The stellar initial mass function (IMF) is a key property of stellar populations.
There is growing evidence that the classical star-formation mechanism by the direct cloud
fragmentation process has difficulties reproducing the observed abundance and binary properties
of brown dwarfs and very-low-mass stars. In particular, recent analytical derivations
of the stellar IMF exhibit a deficit of brown dwarfs compared to observational data.
Here we derive the residual mass function of brown dwarfs as an empirical
measure of the brown dwarf deficiency in recent star-formation models with respect to observations
and show that it is compatible with the substellar part of the
Thies-Kroupa IMF and the mass function obtained by numerical simulations.
We conclude that the existing models may be further improved by including a
substellar correction term that accounts for additional formation channels
like disk or filament fragmentation. The term ``peripheral fragmentation'' is introduced here for
such additional formation channels.
In addition, we present an updated analytical model of
stellar and substellar binarity. The resulting binary fraction and the
dynamically evolved companion mass-ratio distribution are in good agreement with
observational data on stellar and very-low-mass binaries in the Galactic field, in clusters,
and in dynamically unprocessed groups of stars if all stars form as binaries with stellar companions.
Cautionary notes are given on
the proper analysis of mass functions and the companion mass-ratio distribution
and the interpretation of the results. The existence of accretion disks around young
brown dwarfs does not imply that these form just like stars in direct fragmentation.
\end{abstract}

\maketitle

\section{Introduction}
The stellar initial mass function (IMF) is a key tool for star-formation research
because it mirrors the processes of the formation of stellar populations
\citep{Bas+al:2010,IMF-Review-2013}.
Consequently,
the IMF has been subject to extensive research both observationally and theoretically.
In recent years the majority of the star-formation community has favored the
assumption of continuous star formation from the lowest-mass brown dwarfs (BDs) to the
most massive stars \citep{PaNo02,PaNo04,HenCha08}.
However, there is evidence for at least two separate formation channels for most BDs on the
one hand and most stars on the other. For instance, careful analysis of the observational data reveals a
disagreement between the theoretical predictions of the binary separation distributions
of BDs and stars and the observed ones (\citealt{Bouyetal03, Burgetal03, Maetal03} and \citealt{Cloetal03}).
In this light the key assumption of a uniform or continuous formation mechanism assumed in most star-formation
models needs to be questioned.
Based on observational data and N-body computations, \citet{PaGo2011} conclude that the
birth population of very-low-mass binaries with system masses below 0.2~\tmsun\ must be
very different from that of M-dwarfs.
Another important issue is the ``BD desert''
\citep*{2003IAUS..211..279M}, an observed dearth of BD companions to stars.
Unlike earlier studies like e.g.
by \citet{GreLin06} who interpret this as being related to the companion mass ratio
rather than the absolute mass, 
the more recent survey by \citet{Die+al:2012} deduce
a lower absolute mass limit of companions to stars close to 0.1~\tmsun.

Analytical approaches by \citet[][PN02, from here]{PaNo02}
and \citet[][HC08, and HC09, hereafter]{HenCha08,HenCha09}
deduces the IMF from an analytical description of the distribution of prestellar cloud cores.
While the stellar IMF is reproduced by these approaches, a significant deficit
of BDs and very-low-mass stars (VLMSs, between about 0.08 and 0.2~\tmsun) with respect to
observationally constrained IMF descriptions (\citealt{Kr01,Cha05,TK07,TK08}, the latter three
from here on referred to as C05, TK07, and TK08, respectively, and \citealt{IMF-Review-2013})
appears if realistic properties of the prestellar clouds are assumed.
The fundamental reason why direct fragmentation of a turbulent
molecular cloud rarely produces BDs is that the formation of a BD requires a high-density gravitationally
self-bound but very low mass fluctuation that cannot draw significant amounts of additional mass from
an accretion reservoir (see also \citealt{AdFa96}).
However, HC08 and HC09 speculate that this deficit
might be solved by a refinement of their models by including turbulent fragmentation descriptions.
A different interpretation is that their model is overall correct but reveals that an additional formation channel
is required to match the observations.
The IMF by C05 is an empirical and almost equivalent update of the IMF by \citet{Cha03}.
Fully hydrodynamical computations of whole star-forming clouds by e.g. \citet{BBB03} and \citet{Bate2009a}
reproduce the formation of BDs largely from fragmenting circumstellar disks, although these simulations
tend to overproduce BDs unless radiative heating is included \citep{Bate2009b,Bate2012}.

In this paper we introduce the residual mass function (RMF) as
a correction term for this BD deficiency in the analytical approaches by
PN02 and HC08 with respect to the observation-based IMFs in C05 and TK07.
In Section~\ref{sec:method} the model is described and the RMF
is defined.
The results are presented in Section~\ref{sec:results}, followed by an analysis of the
companion mass-ratio distribution in the overlap region of these two populations and a
discussion of the \emph{BD desert}
and its claimed consistency with empirically determined IMF models
(\citealt{RegMey2011}), in Section~\ref{sec:discussion}. In addition, the term \emph{peripheral fragmentation}
is introduced as the main formation channel of BDs.

\section{Method}\label{sec:method}
The basic idea is to quantify the deficit of analytically derived mass functions with respect to
observationally constrained IMFs. The resulting deficit defines the RMF.

\subsection{Observational IMF models}\label{ssec:obs}
The stellar IMF is based on an extensive analysis of observational data from young stellar clusters
and in the Galactic field \citep{KTG93,Kr01}.
Because the majority of multiple systems remain unresolved such IMFs need to be
interpreted as system mass functions unless a careful numerical binary correction
is used.

The empirical IMF by C05 used here is a system IMF based on the earlier IMF by \citet{Cha03},
assuming BDs and stars forming the same way, thereby simply occupying different regions in the
same continuous mass function, an assumption also made in the past by \citet{Kr01,Kr02}
TK07, on the other hand, provide an individual-body IMF that has been transformed into the
corresponding system IMF by Monte Carlo random pairing among the \emph{star-like} and the
separate \emph{BD-like} population introduced in TK07.
This accounts for the observational evidence for two separate (albeit related) formation channels.
In particular, there is a lack of observed BD companions
to stars, especially for small separations \citep{2003IAUS..211..279M,GreLin06},
whereas the statistical properties of binary
BDs differ largely from stellar ones \citep{Bouyetal03,Burgetal03,Maetal03,Cloetal03}.
As argued in \citet{TK07}, the different binary properties of BDs and
VLMSs on the one hand, and stars on the other, consequently suggest the existence of two separate
formation channels. This thus leads to the requirement for two separate IMFs,
each corrected for unresolved binaries, to describe the real individual-body mass function
of a star-forming event.

These IMF descriptions, however, are purely observationally motivated and
therefore do not explain the underlying physical processes leading to the actual
mass spectrum.

\subsection{Analytical IMF models}\label{ssec:ana}
\begin{figure}
\begin{center}
\plotone{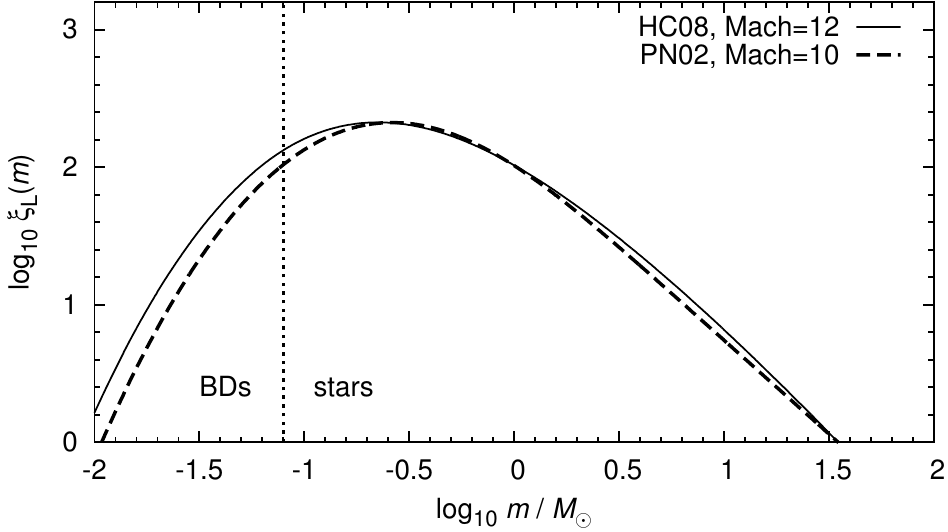}
\caption{\label{hc08-pn02}Comparison of the analytical system IMF by PN02 for ${\cal M}=10$
(their preferred value, dashed curve)
with that by HC08 (${\cal M}=12$, solid curve). Both functions are scaled in this plot for equal peak height
for comparison only.
Although they nearly match in the stellar regime there are slight deviations in the substellar regime
as well as in the positions (i.e. masses) of the peak.}
\end{center}
\end{figure}

In recent years, several attempts have been made to understand the physics
behind star formation and to reproduce its outcome.
This section deals with the analytical models by PN02 and HC08.
In both PN02 and HC08 the prestellar clump mass function
is described by an analytical function with the Jeans mass, the length scale of the
initial inhomogeneities and the Mach number, ${\cal M}$, as the main parameters (see Equation (24) in PN02
and Equation (44) in HC08).
Both studies derive the IMF from the mass distribution of gravitationally unstable clumps
based on empirical power spectra of turbulent flows within the molecular clouds.
While the particular formalism differs in PN02 and HC08, fragmentation due to the
supersonic interaction of gas sheets is the engine for forming Jeans-unstable
clumps in both models.
The resulting prestellar clump mass function is then transformed into a stellar IMF
assuming a star-formation efficiency of 30\pct--50\pct.
The models have been tested in the range $3\le{\cal M} \le 20$.
In this paper, we restrict the model to the Mach number ${\cal M}=12$ in the case of HC08 (Figure 5 top therein)
and ${\cal M}=10$ in PN02 because these Mach numbers are suggested as the most likely values in
these papers (see p. 873 in PN02 and p. 406 in HC08). With these values both models match each other well
in the stellar regime despite their different formulations.
Higher Mach numbers would mainly shift the
mass spectrum toward lower masses and decrease the MF toward the high-mass end.
As shown in Figure~\ref{hc08-pn02} the PN02 ${\cal M}=10$
mass function (long-dashed curve)
and the ${\cal M}=12$ HC08 mass function (solid curve) closely match in the stellar regime but
are slightly different at the low-mass end.
The resulting  mass function deviates
significantly from the observed stellar+substellar mass function in the substellar mass regime,
as shown in Figures \ref{hc08-to-imf} and \ref{pn02-to-imf} for both theoretical models compared
to the observationally constrained IMFs by C05 and TK07.
To quantify this difference the model IMF, either HC08 or PN02, is subtracted from an observational
reference mass function that is taken from C05 and TK07 in order to determine two estimates of
the RMF:
$\xires$,
\begin{equation}\label{eq:rmf}
\xires(m)=\xiobs(m)-\ximod(m)\,,
\end{equation}
where $\ximod$ is any of the theoretical IMFs and $\xiobs$ refers to any of the
observationally constrained IMFs.
In general, a mass function $\xi$ is defined as the differential number $N$
over the differential object mass $m$:
\begin{equation}
\xi(m)=\diff{N}{m}\,,
\end{equation}
and, in the logarithmic scale
\begin{equation}
\xil(m)=\diff{N}{\lgd m/\msun} = (\ln 10)\,m/\msun\,\xi(m)\,.
\end{equation}
The TK07 IMF is the canonical IMF \citep{IMF-Review-2013} that takes into account
that BDs and some VLMSs need to be added as an additional population called
\emph{BD-like}, whereas most stars belong to the \emph{star-like} population:
\begin{equation}
\begin{split}
\label{eq:imf}
\xi_{\rm BD} (m) &= \rpop\,k
  \begin{array}{ll}
   \left(\frac{m}{0.07}\right)^{-0.3}\,, &0.01 < m \simless 0.15,
  \end{array}\\
\xi_{\rm star} (m) &= k\left\{
  \begin{array}{ll}
   \left(\frac{m}{0.07}\right)^{-1.3}\,,  &0.07 < m \le 0.5,\\
   k_m\,\left(\frac{m}{0.5}\right)^{-2.3}\,, &0.5 < m \le \mmax\\
  \end{array}\right.\,,
\end{split}
\end{equation}
where $\rpop$ is the BD-like to star-like population ratio ($\rpop\approx0.2$, TK07) and
$k_m=\left(\frac{0.5}{0.07}\right)^{-1.3}$ ensures continuity of the stellar
IMF at 0.5~\tmsun. Here, $\mmax$ follows from the $\mmax$-$\mecl$ relation
(\citealt{Weidner+al2010,Weidner+al2013}, Equation (10) in \citealt{PWK2007}) and approaches 150~\tmsun\ for the most massive
clusters.
{\it Note the overlapping mass ranges of the populations indicating
that bodies between about 0.07 and 0.15~\tmsun\ may belong either to the star-like
or the BD-like population.}

At the high-mass end of the BD-like population, the sharp truncation
used in TK07 has been replaced in this study by a steep power-law function
to reduce numerical artifacts in the sum IMF. We chose a power-law exponent of
10 to keep the effect on the BD-like to star-like ratio negligibly small. There is no
such declining function applied to the star-like population that is intrinsically smoother
due to the dynamical population synthesis (DPS) method described in Section \ref{ssec:mcbin}.
In addition to this, star-like objects below 0.06~\tmsun\ and BD-like ones
above 0.3~\tmsun\ are not considered. It should be noted here that the theoretical
mass function obtained by \citet{Thiesetal2010} from
smoothed particle hydrodynamics (SPH) simulations also
shows a steep decline above 0.1~\tmsun\ rather than an exact truncation.

Since the C05 stellar IMF used
by HC08 is a system mass function rather than an individual-body-mass
function, the TK07 canonical IMF has also been transformed into
the corresponding TK07 system mass function using the Monte-Carlo model described in
Section \ref{ssec:mcbin}.
The stellar component of the canonical TK07 system IMF is normalized to the
C05 system IMF in the whole mass regime (0.01--150~\tmsun), i.e.
\begin{equation}
\int\limits_{0.01~\msun}^{150~\msun} \xi_{\mathrm{TK07,sys}}(m)\,\dif m=\int\limits_{0.01~\msun}^{150~\msun} \xi_{\mathrm{C05,sys}}(m)\,\dif m\,.
\end{equation}
Here it is assumed that in the C05 IMF stars and BDs
share the same IMF without a discontinuity or overlap, so the mass regime above 0.08~\tmsun\
corresponds to the stellar component in the TK07 model.

The BD multiplicity fraction is only about 10\pct\--20\pct\
(\citealt{Bouyetal03,Cloetal03,Maetal03}; \citealt*{Krausetal06} and
\citealt{LaHoMa08}) and is assumed to be equal for both C05 and TK07
IMFs.

Further, it has to be noted here that the RMF is based
on the mass range between 0.01 and about 0.3~\tmsun\ ($\lgd m/\msun$
between $-2$ and $-0.5$) only, because the residuals in the stellar regime are due to slightly different
functional descriptions rather than being physically constrained and are therefore neglected here.
The residuals between the observational
stellar IMF fits of C05 and TK07 and the HC08 and PN02 models are of potential
interest to later work, but they are not further considered in this paper.

\subsection{Monte-Carlo model of the binarity of stars and brown dwarfs}\label{ssec:mcbin}
Besides the mass function itself, the binarity as a function of the primary-star mass
and the CMRD are also important characteristics of stellar populations.
They are studied here using a Monte Carlo approach with star-like and BD-like objects drawn from separate
IMFs.

The binarity or binary fraction, $\fbin$,
is defined as the ratio of the number of binary or higher-order
systems, $\nbny$, to the total number of systems, $\nsys$. Here, the term \emph{system} includes
multiple systems and singles (their number being noted as $\nsng$) as well. Then
\begin{equation}\label{ftot}
\fbin=\frac{\nbny}{\nsys}=\frac{\nbny}{\nsng+\nbny}\,.
\end{equation}
For the star-like population we apply the binary DPS method developed by
\citet*{MKO11} and \citet{MK11}, hereafter referred to as dynamical or DPS pairing.
In DPS the binary stars are formed in a population of embedded clusters within which they are dynamically
processed to yield the Galactic disk stellar single-plus-binary population. An attractive feature of
this theory is its underlying assumption of the universality of binary properties of late-type stars
being consistent with observational data \citep{MK12,Leigh+al:2014}.
For initial binaries with intermediate to large separations the DPS pairing method applies random
pairing\footnote{Marks et al. (2014, in prep.) show that random pairing with subsequent dynamical processing
does indeed reproduce at least the observed low-mass stellar population (see also \citealt{Kr95b}).}
below a primary mass of 5~\tmsun\ and ordered pairing (such that $q\ge0.9$) above. Here,
$q=\msec/\mpri\le 1$, where $\msec$ is the companion mass and $\mpri$ is the mass of the primary star.
This initial binary population is then altered by dynamical evolution.
Close binaries with orbital periods below about 10 days undergo \emph{eigenevolution} \citep{Kr95b} and tend
to equalize the companion masses.
Note that this \emph{eigenevolution} term alters the very-low-mass end
of the starlike IMF. For the purpose of this work, however these effects only play a negligible role.
Here, the initial or primordial binary fraction is 100\pct, i.e. it is assumed that all stars form in
binaries. The final (after dynamical processing in the embedded cluster)
overall binary fraction is about 40\pct\ (i.e. $\fbin=0.4$) but varies as a function
of the primary-star mass.
For M-dwarfs, in particular, it is as low as 25\pct while G-dwarfs show about 56\pct\ binarity.
The binary fraction approaches 90\pct\ for O stars.
For the BD-like population we chose an overall binary fraction of 20\pct\ (i.e. $\fbin=0.20$),
in accordance with TK07, TK08.
About half of the members of observed average stellar populations are binaries, most of them remaining
unresolved in typical star-cluster surveys. However, very young and likely dynamically unevolved
populations like the Taurus-Auriga association exhibit almost 100\pct\ binarity \citep{IMF-Review-2013,DK2013,
Rei+al2014}.
The number of systems must not be confused with the number of individual bodies, $\nbod=\nsng+2\nbny$.
Since higher-order multiples are relatively rare \citep{GoKr05} they are summarized within the binary population
in this work, so the total number of bodies is
\begin{equation}\label{ntot}
\nbod=2\nbny+\nsng\,.
\end{equation}
The CMRD describes the relative number of binaries as a function of the
companion-to-primary mass ratio.
Observations reveal a continuous decline of $\fbin$ as a function of the primary-object mass which has been
interpreted as a continuous transition from the stellar to the substellar regime
(\citealt{Joergens2008,KraHil2012}, but see \citealt{TK08}).
There is also a shift towards more equal-mass binaries ($q=1$) for VLMS
and BDs \citep{Die+al:2012}. These properties of the stellar population are well reproduced by DPS
such that the origin and properties of binary populations are well understood.

In this study we used a Monte Carlo approach with the TK07 IMF (Equation (\ref{eq:imf})),
i.e., with two separate
populations, \emph{BD-like} and \emph{star-like}, ranging from 0.01 to 0.15~\tmsun\ and
from 0.06 to the maximum stellar mass of 150~\tmsun, respectively. The slightly larger
mass ranges compared to Equation (\ref{eq:imf}) are due to the steep power-law decline added
to the mass borders to reduce numerical artifacts. The masses are
drawn randomly from each IMF where the relative normalization between both populations
is simply obtained by the number of objects drawn from each IMF. 

BD-like objects are assumed to form as single
substellar cores some of which are subsequently paired to binaries within a dense dynamically
preprocessed environment like a massive extended accretion disk
\citep{StHuWi07,Thiesetal2010,BasVor2012}.
Similarly, stellar binaries are assumed to be assembled from individually formed and
subsequently paired stars.
After the individual-body populations have been drawn from the IMF,
two different methods are used for assembling the binaries.
For stars we apply the DPS pairing method mentioned above.

BD binaries, on the other hand, are created by first drawing two objects from their
separate population characterized by $\xibd$ (Equation~\ref{eq:imf}) with
the pairing probability $\ppair$ being determined by a power law,
\begin{equation}\label{qfunction}
\ppair=\mbox{const}\,q^{\fq}\,,
\end{equation}
following the power-law bias scheme used by \citet{Goodwin2013} for the mass ratio distribution
of binaries from second (i.e. binary-forming) fragmentation.
Whereas the case $\fq=0$ corresponds to random pairing from the IMF,
$\fq>0$ describes a biased pairing rule with an increasing pairing probability
toward equal-mass components.
The extreme case of perfect equal-mass pairing corresponds to $\fq=\infty$ but
is in disagreement with data from the Very-Low-Mass Binary Archive (VLMBA; \citealt{BurgetalPPV})
which also contains a few unequal substellar binaries.
In practice, first an array of BD-likes is generated by drawing randomly from $\xibd$
(Equation~\ref{eq:imf}). All subsequent pairings to make binaries are performed on the array only.
Random pairing is performed by randomly drawing a
companion from the array for each object of the same population.
If the companion is more massive than
the considered object it becomes the primary, and otherwise it is the secondary component.
In the case of biased pairing an additional decision is made whether a companion randomly
drawn from the population is accepted to be a partner or rejected,
depending on the probability $\ppair$ in Equation (\ref{qfunction}).
A rejected partner may later be chosen as a partner
to another object. Objects that, on the other hand, are already bound in a binary are
skipped in the pairing procedure henceforth.
The biased pairing procedure is iterated until the required binary fraction $\fbin$ is achieved.
Because the objects are chosen in random order, this does not introduce any additional bias besides
$\ppair$ to the binary populations. The method ensures that the slope of the canonical IMF is retained,
in contrast to methods that select components according to a mass-ratio distribution only.

\section{Results}\label{sec:results}
\subsection{Residual Mass function for Semianalytical Models}\label{ssec:rmf}
\begin{figure}
\begin{center}
\plotone{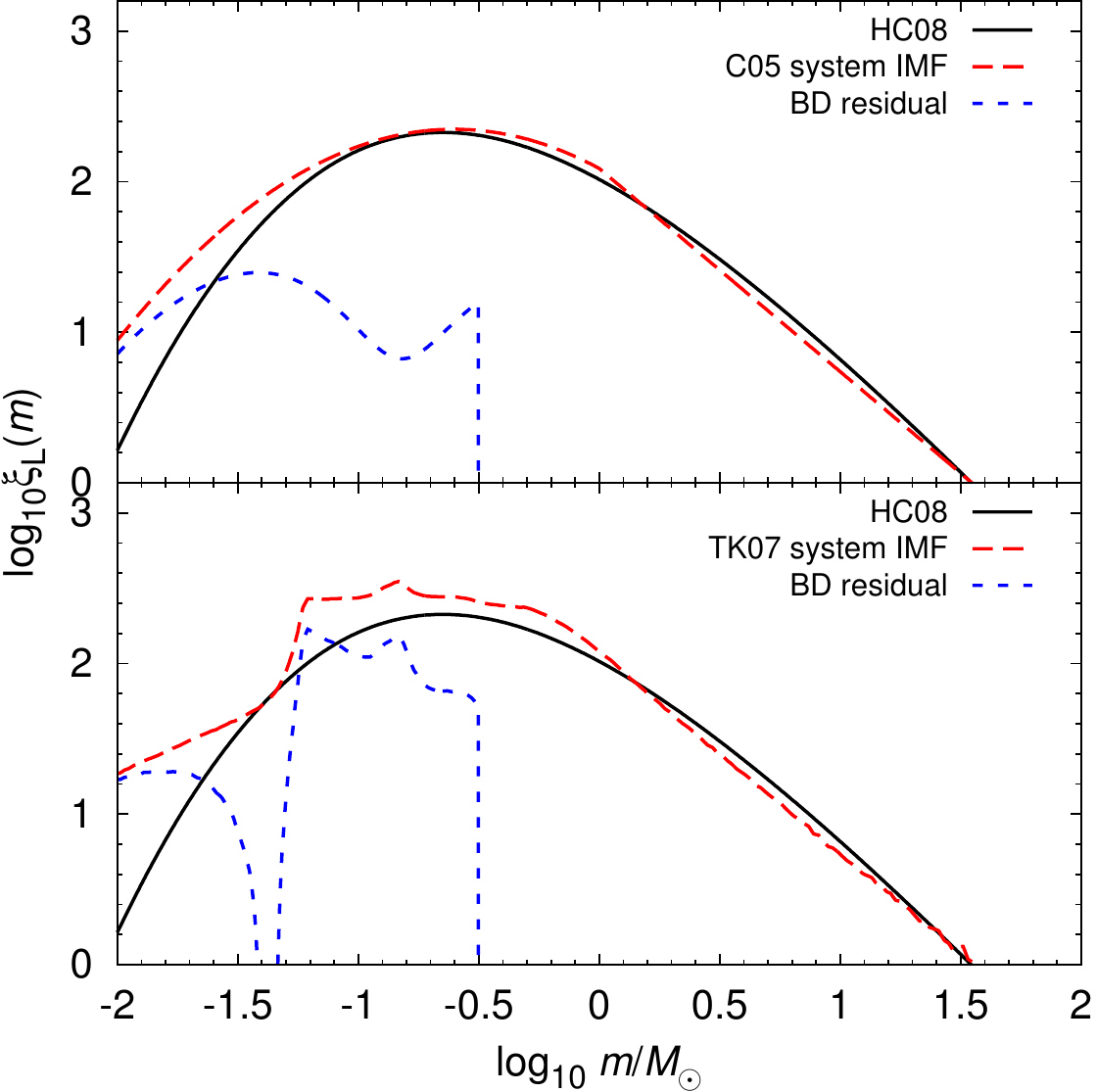}
\caption{\label{hc08-to-imf}{\bf Upper panel:} the analytical IMF model for ${\cal M}=12$ by HC08 (solid line)
compared to the empirical IMF by \citet{Cha05} (dashed line). These functions, originally defined
as system mass functions, have been normalized to match in the stellar regime.
The dotted line represents the residual mass function, i.e.
the difference between both mass functions. The gaps in the RMFs are caused by the intersection of the
HC08 model IMF with the empirical IMFs. They are highly sensitive to the normalization of the IMFs.
{\bf Lower panel:} same as in the upper panel but with the combined bimodal
IMF according to \citet{TK07} (dashed line). Note that the RMF is truncated for $\lgm\ge-0.5$
because the different functional shapes of the stellar parts are not considered here.}
\end{center}
\end{figure}

In this section we present the RMFs obtained from the BD/VLMS deficit
of the analytical IMF models by PN02 and HC08. The RMF is derived for a particular and, according
to PN02 and HC08, typical parameter set, in particular the turbulent Mach number.
Other parameters yield different RMFs, so only the
general functional shape for a likely parameter set is presented here. It is further shown
that such an RMF provides a correction term to the stellar IMFs in order to fit a reasonable
overall IMF without the BD deficit observed in the purely stellar ones.
Figure \ref{hc08-to-imf} shows the results of our calculations
for the analytical model from HC08 with the C05 and TK07 system IMF as the observational reference
(upper and lower panels, respectively).
In both cases the RMF is limited to the mass range below 0.3~\tmsun\ ($\lgm<-0.5$) because the
slightly different shapes of the IMFs in the stellar region are beyond the scope of this paper.
The local minima near $\lgm=-0.8$ (upper panel) and $-1.4$ (lower panel)
occur because the difference between the HC08 mass function and the empirical mass function almost vanishes
locally.
This behavior is highly sensitive to the normalization of the IMFs and is most prominent in the case
of the discontinuous TK07 IMF.
Peaking near the hydrogen-burning
mass limit the RMF declines steeply and effectively vanishes above $\lgm=-0.5$. The peak
mass, however, is better represented in the residual to TK07 in
Figure~\ref{hc08-to-imf} whereas the RMF from C05 appears to be shifted toward
lower masses by about a factor of two (i.e. an offset of about -0.3 on the logarithmic
scale).

\begin{figure}
\begin{center}
\plotone{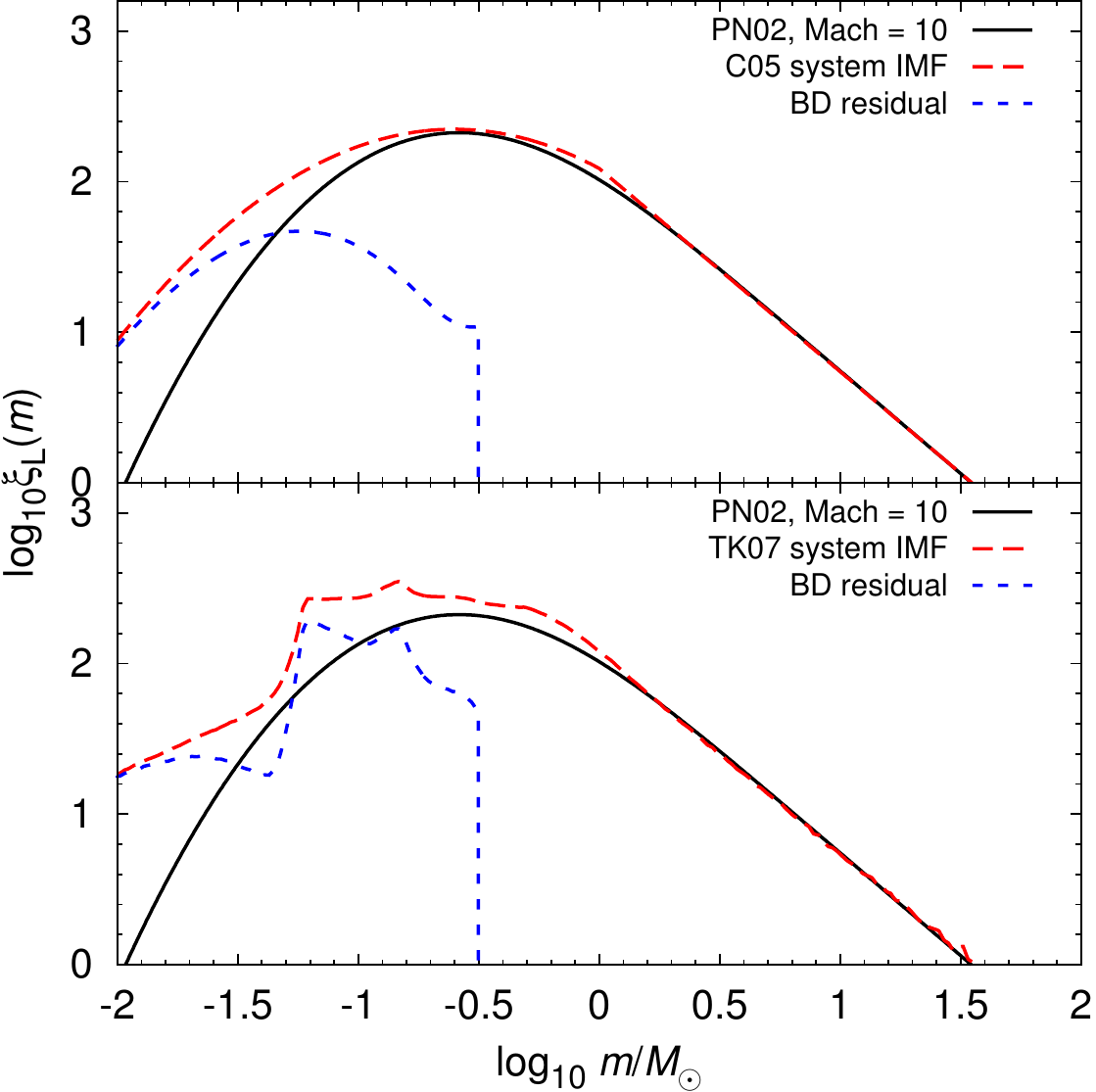}
\caption{\label{pn02-to-imf}{\bf Upper panel:} the same as in Figure~\ref{hc08-to-imf}
but with the clump mass function model by PN02 instead of HC08.
{\bf Lower panel:} same as Figure~\ref{hc08-to-imf}, lower panel, but with the
clump mass function model by PN02 instead of HC08. Here, the RMF does not contain
any gaps.}
\end{center}
\end{figure}

In the case of the PN02 analytical model, as shown in Figure~\ref{pn02-to-imf},
the result with respect to the C05 IMF covers a mass range similar to HC08
but is also continuous.
This is expected because both analytical models
have a functional form similar to the C05 IMF, namely an extended log-normal-type shape.
In the case of the TK07 system IMF (Figure~\ref{pn02-to-imf}) there is a prominent ``dip''
in the RMF around $\lgm=-1.3$ (i.e. $m=0.05\,\msun$).
The reason is the near-equality of TK07 and PN02 at this point, which
varies sensitively with the normalization of the TK07 system IMF and its substellar
component. As with the HC08 model, the RMF of PN02 versus C05 peaks at lower masses
by a factor of about two.

\subsection{Comparison with Simulation Data}\label{ssec:sph}
\begin{figure}
\begin{center}
\plotone{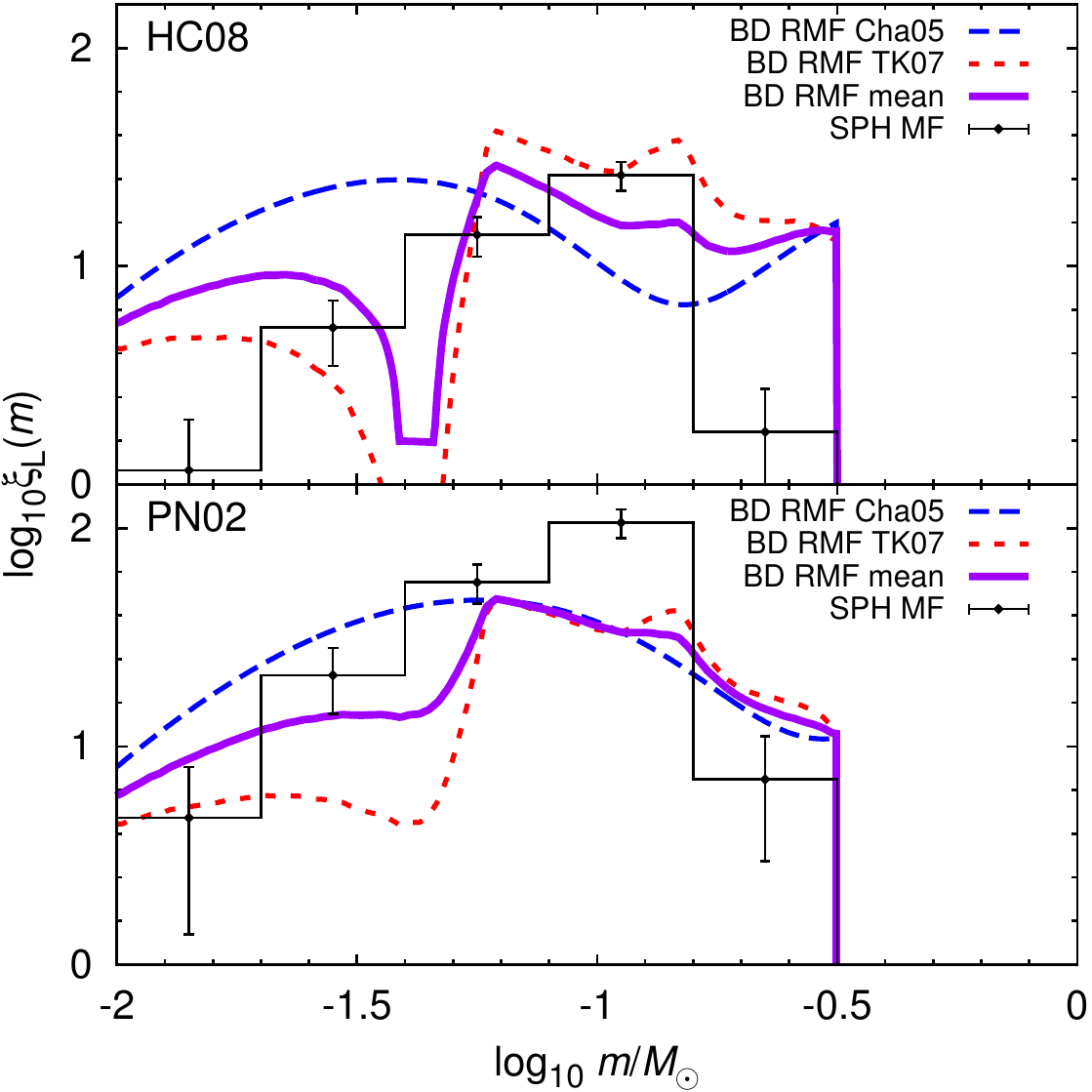}
\caption{\label{rmf-sphmf} Residual mass functions of HC08 (upper panel, from Figure~\ref{hc08-to-imf})
and PN02 (lower panel) with respect to the system mass functions by C05 (dashed line, from Figure~\ref{pn02-to-imf})
and TK07 (dotted line) as well as the average of both (solid line)
in comparison to the SPH mass function from \citet[][including post-population SPH data]{Thiesetal2010},
shown as a histogram (normalized in $\lgd \xi_L$ to the RMF).}
\end{center}
\end{figure}
The RMF obtained in the previous section is compared here with the mass spectrum
of substellar clumps formed in the SPH models of \citet{Thiesetal2010}.
Therein the induced fragmentation of massive extended circumstellar
disks due to perturbation by passing stars in an embedded cluster has been studied. Using an SPH code with
a radiative cooling approximation \citep{Staetal07}
gravitational instabilities have been demonstrated to form through tidal perturbations
and to form
compact clumps with typical masses between 0.01 and 0.15~\tmsun.
The mass function
of the thus-formed clumps, the ``SPH mass function'' (SPH MF) hereafter,
has been shown in \citet{Thiesetal2010} to be in agreement with the substellar component
of the observed system IMF from TK07. In the current study, the SPH mass function
is based on 80 objects formed in 29 computations, some of them performed in continuation of TK07.

In Figure~\ref{rmf-sphmf} the (scaled) SPH mass function is compared to the corresponding RMF.
The overall shape of the RMF essentially matches
the SPH mass function which features a general increase with
increasing mass, a peak around 0.1~\tmsun\ and a rapid decay for higher masses, thus characterizing
a population that is restricted to the substellar and very-low-mass stellar regime.
As mentioned by \citet{Thiesetal2010}, these results are quite similar to those in
\citet{StaWhi09a} who computed BD formation in self-fragmenting disks. Therefore, the results from
triggered fragmentation can be taken as representative for the general disk fragmentation
outcome. Both scenarios assume that the substellar clumps formed in a fragmenting disk
are ejected by either mutual dynamical interactions of clumps in the disk
or in subsequent stellar encounters.
According to \citet{BasVor2012} some of these clumps are disrupted by the tidal forces
during the ejection or are destroyed because of migration onto the central star.
The impact of this effect on the mass distribution will be subject to future
research.

Given the uncertainties of the PN02 and HC08 MFs the RMF based
on them can be considered to be in agreement with the substellar
IMF in both TK07 and \citet{Thiesetal2010}. Interestingly, the average of both
RMFs (each weighted as 50\pct) provides an even better fit to the SPH data (see the solid
curves in Figure~\ref{rmf-sphmf}). Use of such an average may be motivated by the expectation
that the true IMF below $\approx 0.2\,\msun$ is likely between the TK07 and C05
quantifications.

\subsection{Monte Carlo Study Applied to Observations}\label{ssec:comp-imf}
\begin{figure}
\begin{center}
\plotone{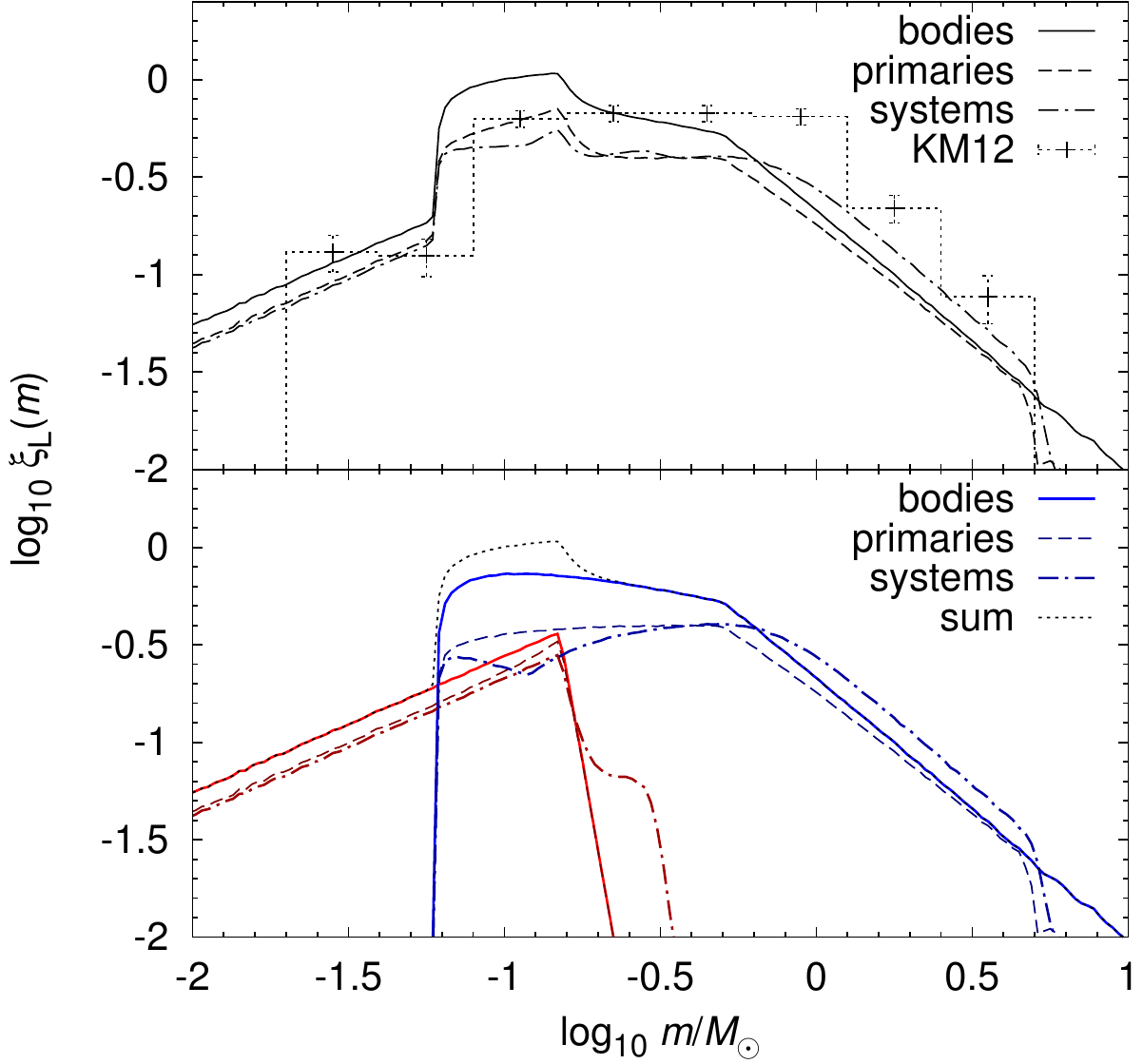}
\caption{\label{split2imf}{\bf Upper panel:} the combined IMF of BD-like and star-like objects from our
Monte Carlo calculations for individual-body masses (solid curve), primary-body masses (dashed curve),
and system masses (dash-dotted curve). The BD-like binarity is 20\pct\ here, and the star-like one is
here about 80\pct\ on average, to match the case of dynamically unevolved populations (see text).
For comparison, the histogram for grouped stars in \citet[][``KM12'']{KM2012}, which also
shows a discontinuity near the stellar--substellar border, is included as
the dotted histogram.
{\bf Lower panel:} the IMFs for the BD-like and star-like
populations plotted separately. The line patterns are the same as in the upper panel for each
separate IMF. For comparison, the individual-body sum IMF is superimposed (thin dotted curve,
the same as the solid curve in the top panel). Note that the very-low-mass end of the star-like
IMF is slightly depleted due to \emph{eigenevolution} \citep{MKO11}.}
\end{center}
\end{figure}

The IMF from the Monte Carlo study described in Section \ref{ssec:mcbin}
is shown in Figure~\ref{split2imf} in comparison to the system mass function
(dash-dotted curve), the primary-body mass function (dashed curve) and the individual-body mass function
(solid curve).
The upper panel displays the sum IMF with both populations combined, as it would be
seen by an observer. In the lower panel, both populations are plotted separately.

We note that Figure 12 in \citet{KM2012}, which shows
the observed average mass functions of young stellar clusters, groups and isolated
objects across the stellar--substellar boundary, reveals a prominent step near the stellar--substellar
border that is very similar to the one that emerges from our analysis as plotted
in the upper panel of Figure~\ref{split2imf}. The system MF
(dash-dotted curve) is to be compared with the mass function by \citet{KM2012}.
Because \citet{KM2012} only studied low-density, dynamically unevolved populations, the
stellar system IMF is modelled here with a binary fraction of 80\pct\ (see also fig. 10 in
\citealt{MKO11}). For the Galactic field, for which the rest of our study has been performed,
 the binary fraction is only about 40--50\pct\ for low-mass stars.

Figure~\ref{split2imf-f} depicts as the solid line the binarity as a function of the
primary-body mass (i.e. the system mass in the case of single objects) from
the DPS pairing model for the Galactic field population
for embedded clusters with a half-mass radius of 0.1~pc for the underlying cluster population.
The long-dashed line indicates the case of a half-mass radius of 0.2~pc which yields
an up to 20\pct\ higher multiplicity. These model Galactic field populations, computed with DPS
(Section \ref{ssec:mcbin}), assume that all stars form in binary systems in a population of embedded
star clusters that evolve and disperse their stars into the field.

These two cases are not to be taken as statistically strict
confidence limits but rather as two likely cases that have been studied in \citet{MK11},
and the initial stellar binary population is indicated by the shaded region at the top.

The stellar binary fraction resulting from DPS is about $\fbin=0.4$
on average in the low-mass region below 1~\tmsun\ for the case of 0.1~pc.
The single-hatched region indicates the large uncertainty of the observed binarity in the substellar region.
For comparison, Galactic-field observational data are overplotted for A stars (VAST survey, \citealt{DeRosa+al2014}, asterisks),
G and late-F stars (\citealt{DuqMay91}, filled squares), M to G stars (\citealt{Ketal03}; open squares),
M dwarfs (\citealt{FisMar92}; upright triangle), and L dwarfs (\citealt{Reidetal08}; upside-down triangle).

Despite consisting of two separate populations (BD-like and star-like)
the binary fraction declines \emph{continuously}
from the stellar toward the substellar regime. As discussed in \citet{MK11} and \citet{MK12},
this is primarily due to the dynamical processing of the star-like population. In part, the mass
overlap of the star-like and the BD-like population also contributes to this transition. It remains
to be shown in future work that the apparent observed trend toward narrow binary separation distributions
for late-type M dwarfs reported by \citet{Janson+al2014} can also be reproduced with DPS by this overlap.

\subsection{Binarity and Companion Mass-ratio Distribution}\label{ssec:fq}
\begin{figure*}
\begin{center}
\plotone{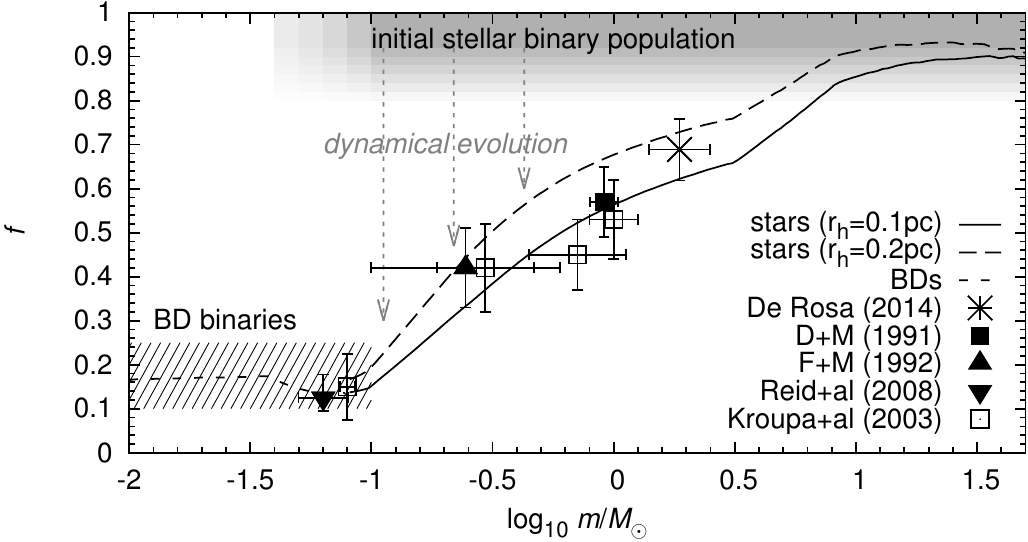}
\caption{\label{split2imf-f}
Binary fraction for BDs and stars as a function of
the primary-object mass. The solid line shows the result of our Monte Carlo computation
for stars via DPS pairing for the field population based on the integration of the star-formation
outcome over all cluster masses. A half-mass radius $\rhm=0.1$~pc is assumed for all embedded host clusters.
The average field-star binary fraction is $\fbin\approx0.5$ for low-mass to intermediate-mass stars, in accordance with TK07.
The long-dashed line refers to the DPS model with $\rhm=0.2$~pc, corresponding to a higher binary fraction.
The short-dashed curve represents the BD-like population with biased pairing. The single-hatched region in the lower
left corner indicates the uncertainty of the substellar binary fraction in the observational data.
The smoothly shaded area at the top refers to the initial stellar binary population with a binary fraction
of near 100\pct\ \citep{MK11}.
Observational field-star data are overplotted for A stars (VAST survey, \citealt{DeRosa+al2014}, asterisks),
G and late-F stars (\citealt{DuqMay91}, filled squares), M to G stars (\citealt{Ketal03}; open squares),
M dwarfs (\citealt{FisMar92}; upright triangle), and L dwarfs (\citealt{Reidetal08}; upside-down triangle).
}
\end{center}
\end{figure*}

\begin{figure}
\begin{center}
\plotone{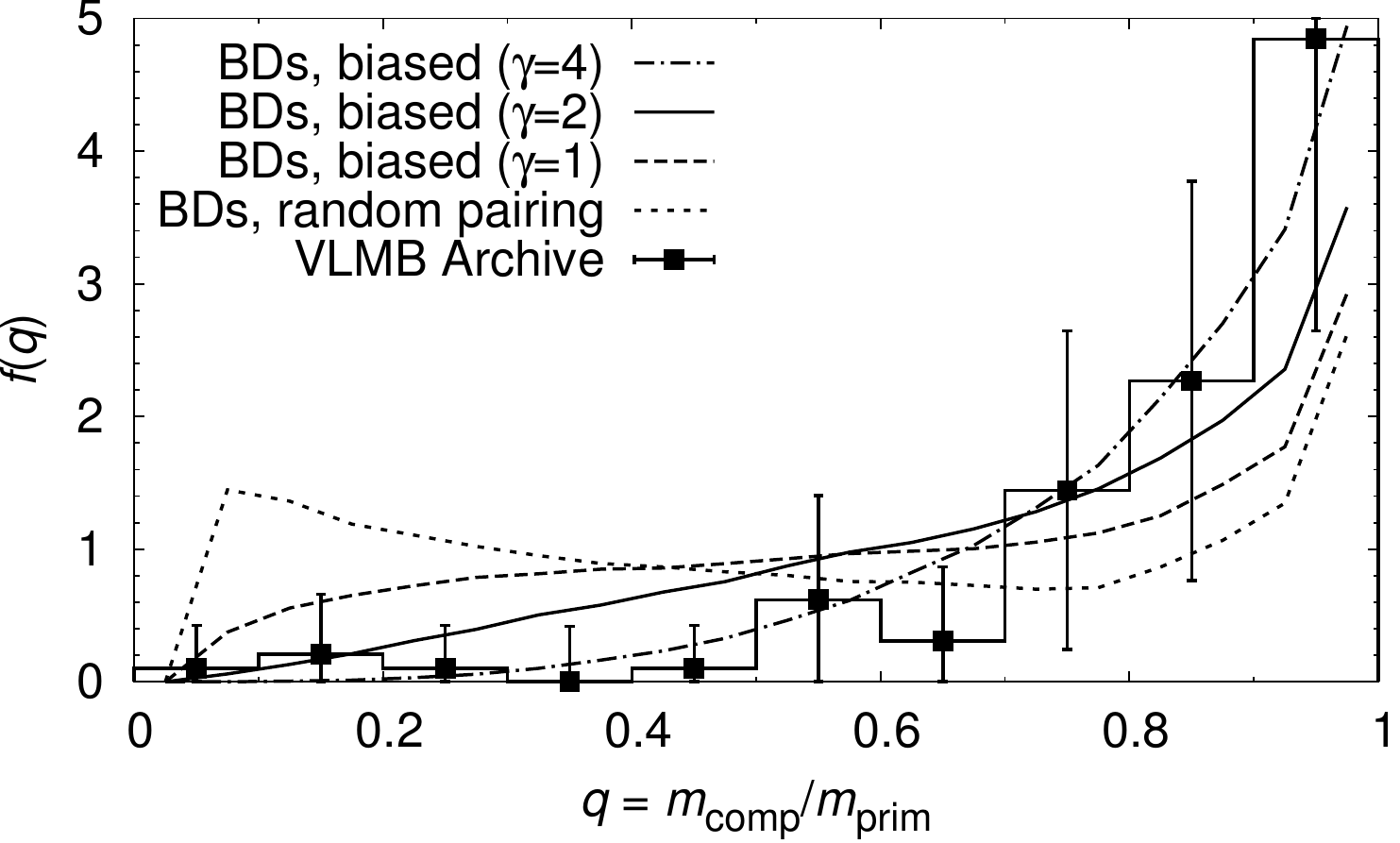}
\caption{\label{split2imf-q}
Number distribution $f(q)$ for binary BDs between 0.03 and
0.08~\tmsun\ as a function of the companion-to-primary mass ratio.
The solid line shows the result of our computation for $\fq=2$, and
the dashed and dotted lines refer to the case $\fq=1$ and random pairing,
respectively. For completeness, also the more extreme case of $\fq=4$ \citep{DK2013}
is also shown as the dash-dotted line.
The histogram shows the observational data from the Very-Low-Mass Binary Archive
\citep{BurgetalPPV}.
The peak at $q=1$ is primarily due to the contribution of binaries from the low-mass end
of the star-like population because any binary with both components near the
lower mass border of a population can only have nearly equal component masses.
The area below each curve and below the histogram
is normalized to 1.}
\end{center}
\end{figure}

The observable BD CMRD,
which is the CMRD for all primaries (BD-like and star-like) below 0.08~\tmsun,
is shown in Figure~\ref{split2imf-q}.
As can be seen the case of strongly biased pairing with $\fq=2$
(solid line) matches the measured data from the VLMBA better (and also see
\citealt{Die+al:2012}) than the
random-pairing $q$ ($\fq=0$, dotted line) and weakly biased pairing models ($\fq=1$, dashed line).
In addition, the case $\fq=4$ (\citealt{DK2013}; dash-dotted line) is shown.
The binarity of BDs near the hydrogen-burning mass limit of 0.08~\tmsun\ is quite low in the model, only about
11\pct--12\pct\ which is largely due to the ``dip'' in the binarity in the VLMS region of the DPS pairing
method (see Figure~\ref{split2imf-f}). This local minimum is a direct consequence of the two overlapping
populations. \citet{Reidetal08} indeed report a similar binary fraction
of $12.5\pma{5}{3}$\pct\ for L dwarfs. If a local minimum is confirmed by future surveys with
sufficient precision in mass this would further support an overlap of two separate populations
in this mass region. However, in the current data this feature is within the uncertainties
of the observational data.

\begin{figure}
\begin{center}
\plotone{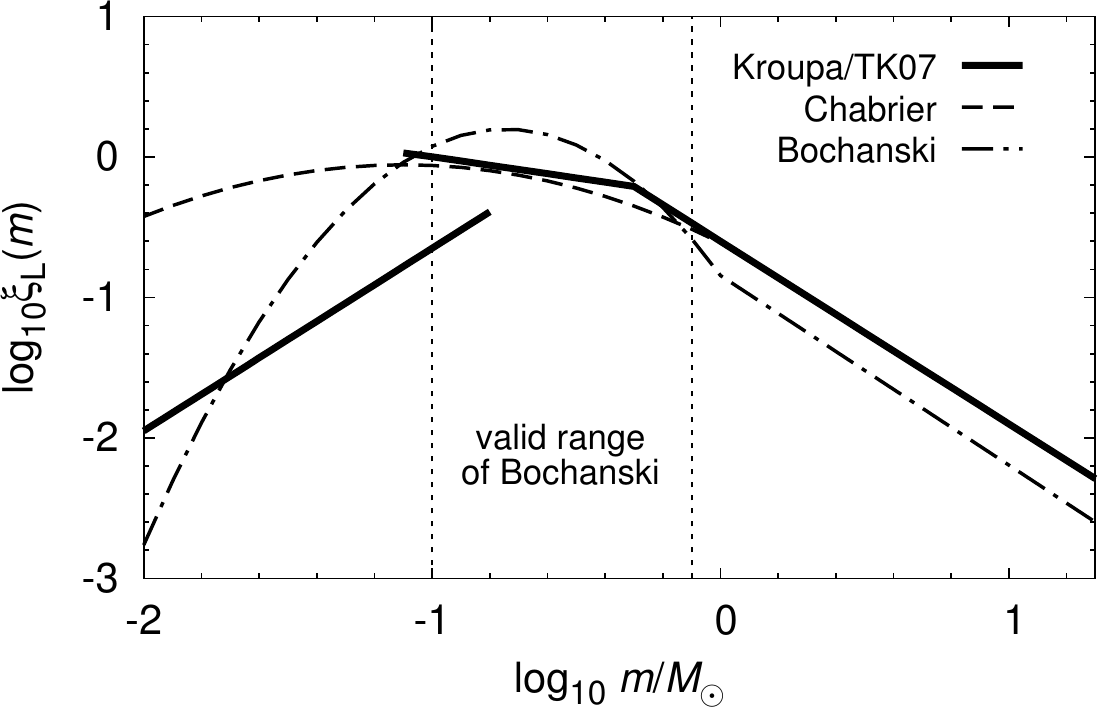}
\caption{\label{imfs-kcb}Comparison of the \citet{TK07}, \citet{Cha05}, and the extrapolated
\citet{bochanski+al:2010} IMF, the latter being used by \citet{RegMey2011}.
The original Bochanski IMF has been derived from data between 0.1 and 0.8~\tmsun.
Note the steeper decline in the Bochanski IMF below its valid mass range relative to the
Chabrier IMF.}
\end{center}
\end{figure}

\subsection{Contribution of Peripheral Fragmentation to BDs}\label{ssec:perifrag}
One important implication of the model discussed in this paper is the hybrid nature of BDs and VLMSs.
Although there is a minor contribution by the star-like population from direct fragmentation, the majority of
BDs are contributed by the BD-like population through dynamically preprocessed gas
(e.g. fragmenting circumstellar disks \citep{StaWhi09a, Thiesetal2010}.
Because this preprocessed material often occurs in the peripheral
regions of star formation (e.g. in the outer parts of disks) we propose the term \emph{peripheral fragmentation}
for the BD-like formation channel.
Of all BDs between 0.01 and 0.08~\tmsun\ 64\pct\ are contributed by the BD-like population,
and 19\pct\ of M dwarfs between 0.08 and 0.45~\tmsun\ are BD-like.
This result, however, is highly sensitive to the chosen lower mass border of the star-like regime. Here,
we assumed it to be 0.06~\tmsun. If, on the other hand, a sharp
truncation of the star-like regime at 0.08~\tmsun\ is chosen, the BD-like fraction of BDs is 100\pct\
but still 18\pct\ of M dwarfs are BD-like.
Because star-like BDs and VLMSs may be detectable by larger circumsubstellar disks or, if applicable, wide binary
separations, future high-resolution observations will help to further constrain the BD-like and star-like mass
borders.

\section{Discussion}\label{sec:discussion}
\begin{figure}
\begin{center}
\plotone{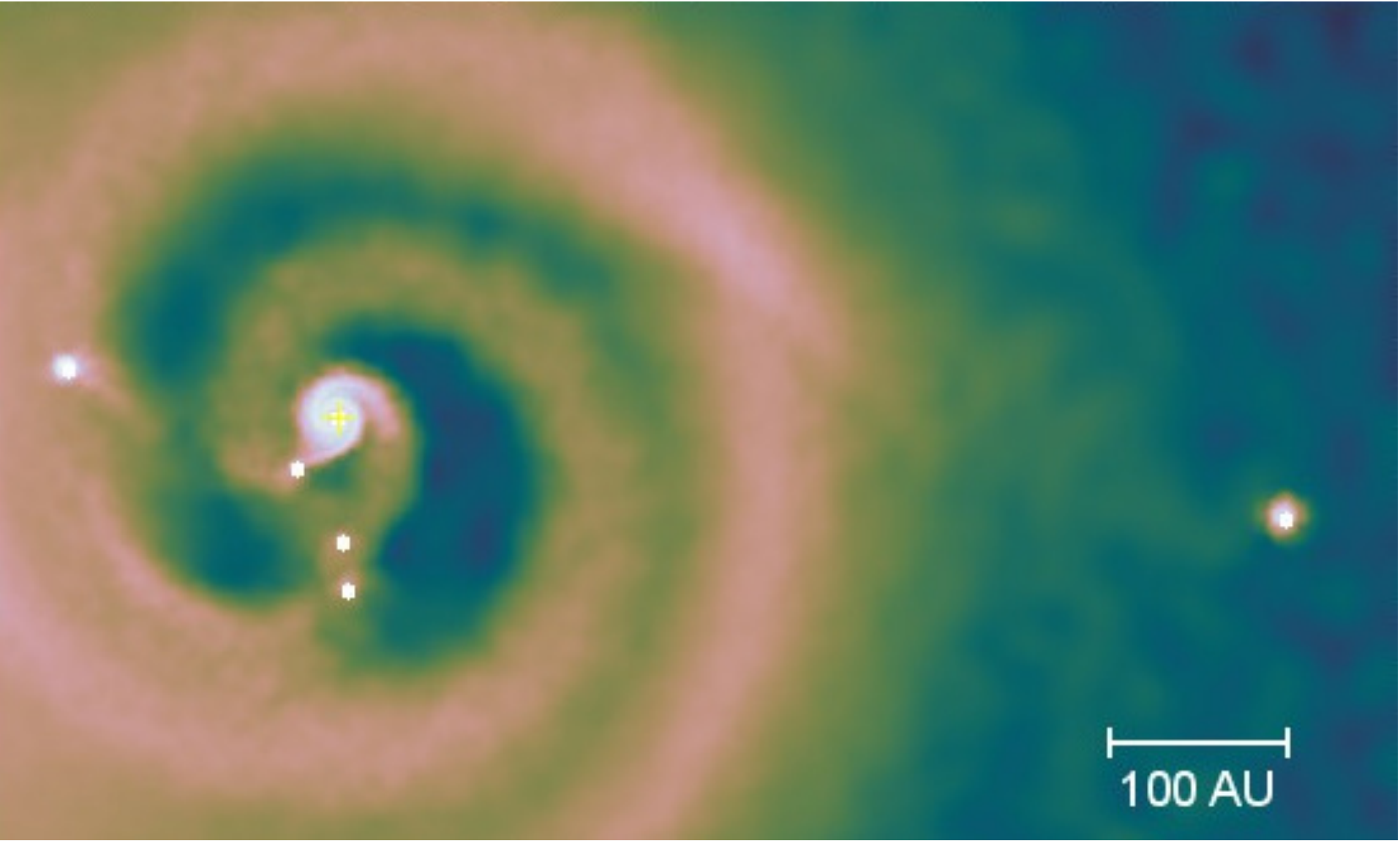}
\caption{\label{imgX002}Snapshot from SPH model X002 by \citet{Thiesetal2010} showing
substellar clumps (with sink particles marked with white points) around a fragmented accretion disk,
all of them surrounded by their own
accretion envelopes. Note that the escaping object (0.013~\tmsun, i.e. a low-mass BD) to the right
has retained its accretion disk (visible as diffuse
structure around the sink particle) even after dynamical ejection. The half-mass radius of the envelope
is about 10~AU, and its mass is \mbox{$2\cdot10^{-4}\,\msun$}, i.e. almost 2\pct\ of the mas of the escaping object.}
\end{center}
\end{figure}

In this paper we have introduced the RMF
as a correction term for the analytical star-formation models by PN02 and HC08 to match the
observationally constrained IMF by \citet{Cha05} and \citet{TK07}.
The effective deficit of BDs and VLMSs in these theoretical models
with respect to the observed BD statistics suggests the requirement for additional
formation channels in these analytical IMF models. HC08 speculate that
this deficit, being at least at the edge of
significance (see uncertainties shown in Figure 4--23 in \citealt{IMF-Review-2013}),
may be solvable by an improved algorithm that accounts for the
effects of turbulence and other dynamical processing of the prestellar gas.
The expected contribution of these additional effects to the IMF may be
understood as a separate population or, at least, as an additional formation channel
of BDs and VLMSs and can formally be described by an additional correction term.
It is also applicable as a test for future star-formation models.

This correction term has been identified and quantified here as the
RMF which has a similar shape for both analyzed observational reference IMFs by C05 and TK07.
The most striking outcome is a general agreement between the RMF (Equation (\ref{eq:rmf}))
as obtained from the theoretical IMF and the observational IMFs, and
the mass function of fragments from SPH simulations \citep{Thiesetal2010}.
In other words, both PN02 and HC08 IMF models
appear to describe essentially the population of stars without a significant fraction
of BDs. They can be considered to be successful models of the direct fragmentation
process in molecular clouds. These models cannot, by their nature, accommodate the peripheral
fragmentation, e.g., in accretion disks around protostars, which yields the BD-like population
(see Section \ref{ssec:perifrag}).
The most natural way to explain these results is, to the best of our knowledge,
a composite population consisting of a star-like and a BD-like component such that
a significant fraction of the BD part comes from an additional process acting
during star formation \citep{ReiCla01,StHuWi07,Thiesetal2010,BasVor2012}.
Essentially this result follows for BDs being less likely to form directly in a star-forming
molecular cloud than from dynamically preprocessed material.
This is because in order to form only a BD a cloud core needs to have a
high pre-collapse density (for it to be gravitationally unstable) while having no further supply
of gas in order to limit its further growth in mass. This has already been emphasized
by \citet{AdFa96}. While there has been a recent discovery
of a possible proto-BD in Oph B-11 with a mass of 0.03~\tmsun\ \citep{Andre+al2012} its further
evolution and possible additional accretion from the surrounding molecular gas remains unclear.
And, the existence of the BD-like population formed through peripheral fragmentation does not
exclude the formation of BDs via the direct fragmentation channel, i.e. as star-like objects
as is evident from the analytical PN02 and HC08 IMFs.

\subsection{Is There a Separate Substellar Population?}
There is an ongoing discussion whether a separate population is really needed to explain
the observed properties.
A similar discontinuity between planets and BDs is widely accepted.
\citet{Cha+al:PPVI} even report a possible mass overlap
of the BD and the giant planet regime as well and suggest a distinction
between BDs and planets by their formation history rather than by deuterium burning.
This can be seen as an analog to the separate substellar population addressed by
our work.
\citet{JumFis2013} claim to be able to model
the trend toward more tightly bound binaries in the BD regime. However, because this study
already assumes the masses to be drawn from a continuous IMF it a priori
excludes the option of a two-component IMF.
The AstraLux survey reports a narrower
binary separation distribution for M-dwarf binaries compared to binaries with FGK primaries
(\citealt{Janson+al2012} and \citealt{Janson+al2014}, respectively).
Such a narrow peak in the separation distribution
may be related to sensitivity artifacts as well as, at least for late-type M dwarfs, due to the
mass overlap with the BD-like population that extends up to about 0.2~\tmsun.

In contrast, in a study of dynamical binary evolution
\citet{PaGo2011} found that the field populations of M dwarfs and very-low-mass binaries
must reflect very different birth populations, since their dynamical processing is essentially
the same.

\subsection{The Brown Dwarf Desert}
Similarly, the BD desert itself has been disputed by several groups who
suggest that it is more related
to the companion mass ratio rather than to the absolute mass.
However, a recent survey by \citet{Die+al:2012} further
supports the existence of a true BD desert,
i.e. a dearth of substellar companions to primary stars ranging from M to G dwarfs.
They found an effective lower mass limit of stellar companions near 0.1~\tmsun\
which cannot be related to incompleteness but rather points toward a discontinuity
in the pairing statistics close to the hydrogen-burning mass limit. They particularly
point out that this discontinuity is essentially independent of the primary star mass.
This clearly supports a separate substellar population contributing the majority of BDs
and is also in agreement with SPH studies of whole star-forming clouds by, e.g.,
\citet{BBB03} and \citet{Bate2009a} who
reproduce the formation of BDs largely from fragmenting circumstellar disks. They, however,
point out that these simulations tend to overproduce BDs and attribute this
to the lack of radiative heating in their model \citep{Bate2009b,Bate2012}.
It should also be noted that \citet{Bate2012} could reproduce a mass function in good
agreement with the C05 IMF and that the previous overproduction of BDs was avoided there
because the radiative heating reduced the frequency of disk fragmentation.

In a related Monte Carlo study performed here, we also found a good
agreement of the two-population composite model with observational data on
field very-low-mass stellar and BD binaries (Figure~\ref{split2imf-f}). The binarity is well reproduced
down to the stellar--substellar border as a continuous function of the primary-star
mass in agreement with the observations. More importantly, the
mass-ratio distribution of very-low-mass binaries deduced from observational
data is well reproduced in our model (Figure~\ref{split2imf-q}), once more supporting a bimodal
star and BD formation scenario. Whereas star-like binaries are well
represented by DPS \citep{MPK15}, the BD-like binaries
apparently do fit a
biased pairing where the probability of pairing rises toward
more equal component masses following Equation (\ref{qfunction}). Here, the
cases $\fq=0$ (simple random pairing) and $1\le\fq\le2$ (biased toward
equal-mass binaries) were compared. The biased pairing case is in
better agreement with the observational data. It remains to be
studied whether postformation dynamical processing through stellar-dynamical
encounters in their birth embedded clusters is responsible
for this apparent preference of more equal-mass binaries over more
unequal ones in the BD mass range.
However, both simple and biased pairing are
in reasonable agreement within the uncertainties of the data if a low
binary fraction of 10\pct\ within the BD-like population is
assumed.
A binary fraction of less than 10\pct\ in the theoretical disk fragmentation outcome
and the very narrow range of semimajor axes
have also been found by \citet{Thiesetal2010}, as well as the survival of circumsubstellar
accretion disks, as shown in Figure~\ref{imgX002}.
The overall binary fraction for systems within 0.03 and
0.08~\tmsun, irrespective of the population to which they belong, is
around 20 \pct\ for both simple and biased pairing (i.e. the value chosen for
the BD-like population), if the local minimum in the binarity
function near the stellar--substellar boundary (Figure~\ref{split2imf-f})
is assumed to be an artifact, whereas it is as low as 11\pct\
otherwise. However, observations of VLMSs indicate a binary fraction as low as 12.5\pct\
\citep{Reidetal08}, so this ``dip'' may indeed be real and thus be a result of the
overlap of two separate populations.

\subsection{Companion Mass Ratio Distribution}
There is an ongoing discussion whether random pairing is
applicable for stars.
\citet{RegMey2011} claim to rule out this model
in favor of a universal CMRD. However, their results only cover a
narrow range of validity of the underlying model IMF (as specified by \citealt{bochanski+al:2010},
see Figure~\ref{imfs-kcb}) that only covers M, K and late-G stars, but no BDs.
Consequently, the CMRD derived from it is quite limited, especially for M dwarfs, for which
no hypothesis test on the pairing rule can be made on this basis. \citet{RegMey2011}
perform pairing experiments that actually exceed the range of validity of the underlying
mass function by \citet{bochanski+al:2010}, and therefore their results are not applicable.
\citet{Goodwin2013} suggests that a random partition of protostellar cloud fragments
is in better agreement with observations of stellar binarity
than an initial binary population drawn from the IMF with subsequent dynamical and eigenevolution.
Because that study was based on the IMF by \citet{Cha03} without a
separate BD treatment
there cannot by any satisfactory agreement with the observed
substellar CMRD. Even more importantly, the initial binary population
needs to be modeled and tested against dynamically barely evolved stellar populations
like Taurus-Auriga rather than against the Galactic field.
The field population then results from an integration
of all dynamically evolved clustered populations following the embedded cluster-mass function
\citep{MK11,MPK15}.

As a final remark, it is being argued that the existence of disks around young BDs implies that these
form like stars supposedly supporting the continuous IMF scenario. Figure~\ref{imgX002} is one
of many examples where a young BD with an accretion disk formed from peripheral fragmentation
in a circumstellar disk is nudged away to become a free-floating BD with a disk.

\section{Conclusion}\label{sec:conclusion}
As the main conclusion from the first part of this study we emphasize
the difficulties of theoretical star-formation models to describe both
stars and BDs by a single mechanism, namely from direct cloud
fragmentation. It remains necessary to treat BDs separately which
implies a separate albeit related formation channel for the majority
of BDs. This already follows from the work of \citet{AdFa96}. The theoretical
evidence by \citet{StaWhi09a,Thiesetal2010}, and \citet{BasVor2012}
supports a separate
population by fragmentation of extended young circumstellar disks.
If O stars are close by, the accretion envelope of VLM protostars may be
photoevaporated, leaving an unfinished substellar embryo \citep{KB03b}.
In addition, the embryo-ejection model by \citet{ReiCla01} gives an
example of BD formation by ejection of unfinished stellar embryos out
of multiple protostar systems. Because these mechanisms are not covered
by the analytical cloud fragmentation models by PN02 and HC08 they do
not contribute to the resulting theoretical clump mass function. The
same is true for BD formation in dense gaseous filaments that form
due to the gravitational pull of surrounding stars \citep{BCB08}
rather than being described by random density fluctuations assumed by
PN02 and C05. Therefore, an analytical model also covering BDs and VLMSs
must include such separate formation channels via dynamically preprocessed
material (disks, unstable multiple stellar embryo systems, photoevaporation,
tidally shaped gas filaments and so on).
The normalization may be adjusted empirically or
deduced from physical relations based on typical disk-to-stellar mass
relations and by the application of turbulence and the Jeans criterion
within this preprocessed material. The development of such an
analytical or semianalytical model should be the subject of future studies.
The observational data thus very strongly suggest that
the dominant fraction (Section \ref{ssec:perifrag}) of
BDs and VLMSs results from peripheral fragmentation that contributes
a separate IMF from that of stars.

This project has been funded by DFG project KR1635/27.

\end{document}